\documentclass[hyper,12pt,letterpaper]{JHEP3}
\usepackage{epic,eepic}
\usepackage{amsmath,epsfig}
\usepackage[latin1]{inputenc}
\usepackage{amssymb,amsfonts}
\usepackage{graphicx}

\title{Spectrum of CHL Dyons from Genus-Two Partition Function}

\preprint{\hepth{yymmnnn}\\TIFR/TH/06-33\\HUTP-06/A0043}
\author{Atish Dabholkar\footnote{Email:
atish@theory.tifr.res.in}$^{~1}$ and Davide Gaiotto\footnote{Email:
dgaiotto@fas.harvard.edu}$^{~2}$\\
\it $^1$Tata Institute of Fundamental Research,\\
\it Homi Bhabha Rd, Mumbai 400 005, India\\

\it $^2$Jefferson Labs,\\
\it 17 Oxford St, Harvard University, \\
\it Cambridge, MA 02138  USA}

\abstract{We compute the genus-two chiral partition function of the
left-moving heterotic string for a $\mathbb{Z}_2$ CHL orbifold. The
required twisted determinants can be evaluated explicitly in terms
of the untwisted determinants and theta functions using orbifold
techniques. The dependence on Prym periods cancels neatly once
summation over odd charges is properly taken into account. The
resulting partition function is a Siegel modular form of level two
and precisely equals recently proposed dyon partition function for
this model. This result provides an independent weak coupling
derivation of the dyon partition function using the M-theory lift of
string webs representing the dyons. We discuss generalization of
this technique to general $\mathbb{Z}_N$ orbifolds.}


\renewcommand{\th}{\theta}

\def\h{\eta}

\def\G{\Gamma}

\def\wp{{\cal P}}

\def\CN{{\cal N}}
\def\IC{\relax\hbox{$\inbar\kern-.3em{\rm C}$}}

\def\IC{{\bf C}}

\def\CN{{\cal N}}

\def\bea{\begin{eqnarray}}
\def\eea{\end{eqnarray}}
\def\be{\begin{equation}}
\def\ee{\end{equation}}
\def\ba{\begin{align}}
\def\ea{\end{align}}
\def\bse{\begin{subequations}}
\def\ese{\end{subequations}}
\def\1F1{{}_1\!F_1}
\def\2F0{{}_2\!F_0}

\def\G{\Gamma}

\def\h3{$\textrm{H}_3^+$}

\def\IC{{\mathbb C}}

\def\lbldef#1#2{\expandafter\gdef\csname #1\endcsname {#2}}

\def\href#1#2{#2}

\newcommand{\beq}{\begin{equation}}
\newcommand{\eeq}{\end{equation}}
\newcommand{\ber}{\begin{eqnarray}}
\newcommand{\eer}{\end{eqnarray}}

\def\be{\begin{eqnarray}}
\def\ee{\end{eqnarray}}

\begin{document}


\section{Introduction}

For certain CHL orbifold compactifications to four dimensions with
$\CN =4$ supersymmetry, there exists a proposal for a partition
function that counts the exact degeneracies of dyonic quarter-BPS
states \cite{Dijkgraaf:1996it, Jatkar:2005bh}. The partition
function for a $\mathbb{Z}_N$  orbifold is proportional to the
inverse of a specific Siegel modular form $\Phi'_k$ of weight $k$ of
a  subgroup of  $Sp(2, \mathbb{Z})$. The weight $k$ is related to
the level $N$ by $k= \frac{24}{N+1} - 2$ for $N = 1, 2, 3, 5, 7$.
The resulting dyon degeneracies satisfy a number of nontrivial
consistency checks. In particular, they are integral, duality
invariant, and in agreement with Bekenstein-Hawking-Wald entropy of
the corresponding black holes \cite{Dijkgraaf:1996it,
LopesCardoso:2004xf, Jatkar:2005bh, Dabholkar:2006mm}.

For partition functions that count the perturbative winding-momentum
states or D-brane bound states, there is a systematic weak coupling
derivation using worldsheet or gauge theory techniques. It would be
desirable if the dyon partition function can be derived in a similar
fashion using worldsheet techniques in an appropriate duality frame.
For toroidally compactified heterotic string discussed in
\cite{Dijkgraaf:1996it}, which corresponds to $N=1$, the dyon
partition function equals the inverse of the Siegel modular form
$\Phi_{10}$ which is the well-known Igusa cusp form \cite{Igusa:ig,
Igusa:ig2}. Using the fact that this is precisely the genus-two
chiral partition function of the left-moving heterotic string, and
an M-theory lift of string webs, a weak coupling interpretation of
the dyon partition function in the $N=1$ case  was proposed in
\cite{Gaiotto:2005hc}. For the $\mathbb{Z}_2$ CHL orbifold, it was
observed in \cite{Dabholkar:2006xa} that the relevant Siegel modular
form  $\Phi'_6$ of level two has the right factorization properties
to be interpreted as a chiral twisted partition function of the CHL
orbifold. Motivated by these results, we explicitly compute the
genus-two partition function of the left-moving twisted bosons in
the $\mathbb{Z}_2$ orbifold and show that indeed it is precisely
proportional to the inverse of the Siegel modular form ${\Phi'_6}$
as expected for $N=2$. The procedure easily generalizes to
$\mathbb{Z}_N$ orbifolds.\footnote{For a complementary and
independent weak coupling derivation using 4d-5d lift in Taub-NUT
geometry, see \cite{Gaiotto:2005gf, Shih:2005uc,Dabholkar:2006xa,
David:2006yn, David:2006ji, Dabholkar:2006mm}.}

This paper is organized as follows. In $\S{\ref{Dyon}}$ we review
the proposal for the dyon partition function. In
$\S{\ref{Genustwo}}$ we explain using various dualities why the dyon
partition function is expected to be proportional to the genus two
chiral partition function of the left-moving heterotic string on the
CHL orbifold. In $\S{\ref{Determinants}}$ we treat the
$\mathbb{Z}_2$ case in complete detail and compute the required
twisted determinants to evaluate the partition function in terms of
the partition function of the unorbifolded theory. The partition
function depends on certain additional parameters called Prym
periods but this dependence neatly cancels against the sum over odd
momenta. The final answer precisely equals the inverse of $\Phi'_6$
as expected. In particular, the weight $k=6$ turns out to be
correlated with the order of the orbifold $N=2$ in precisely the
fashion required for agreement with the black hole entropy. We
discuss generalizations to $\mathbb{Z}_N$ orbifolds in
$\S{\ref{ZN}}$ and conclude in $\S{\ref{Comments}}$ with some
comments.

\section{Dyon Partition Function in CHL Orbifolds\label{Dyon}}

Consider heterotic string theory compactified on $ T^4 \times S^1
\times {\tilde S}^1$. In ten dimensions, the gauge group is $E_8
\times E_8$ or $SO(32)$ of rank $16$ and upon compactification there
are additional $12$ $U(1)$ gauge bosons arising from the
Kaluza-Klein reduction of the metric and the 2-form field. The
resulting theory in four dimensions then has a gauge group of rank
$28$ with ${\cal N} =4$ supersymmetry. The  strong-weak coupling
S-duality group of this toroidally compactified theory is $SL(2,
\mathbb{Z})$.

A CHL compactification that has ${\cal N}=4$ supersymmetry but a
gauge group of reduced rank $r <28$ can be constructed as a
$\mathbb{Z}_N$ orbifold of this toroidal compactification
\cite{Chaudhuri:1995fk, Chaudhuri:1995bf, Chaudhuri:1995dj,
Schwarz:1995bj, Vafa:1995gm}. The generator of the $\mathbb{Z}_N$
group is  $\xi T$ where $\xi$ is an order $N$ left-moving twist
symmetry of the $ T^5$ compactified string and $T$ is an order $N$
shift  along the circle factor $ S^1$. Since the twist symmetry
$\xi$ acts nontrivially on the left-moving gauge degrees of freedom,
some of the $28$ massless gauge bosons are projected out.
Furthermore, because of the order $N$ shift in the orbifolding
action, the twisted states have a $1/N$ fractional winding along the
circle and hence all twisted states are massive. The resulting
orbifolded theory in four dimensions thus has rank smaller than 28.
All ${\cal N} =4$ supersymmetries are preserved because $\xi T$ acts
trivially on the right-moving fermions. The S-duality group of a
$\mathbb{Z}_N$ model is the congruence subgroup $\G_1(N)$ of $SL(2,
\mathbb{Z})$  of matrices
\begin{equation}\label{gamma}
   \left(
     \begin{array}{cc}
       a &b \\
       c & d \\
     \end{array}
   \right), \qquad ad-bc =1, \quad c = 0\, \textrm{mod}\, N, \quad
     a = 1\, \textrm{mod}\, N,
\end{equation}
which acts on the electric and magnetic charge vectors as
\begin{equation}\label{charge}
    Q_e \rightarrow a Q_e + b Q_m, \qquad Q_m \rightarrow c Q_e + d
    Q_m.
\end{equation}
 The
restriction on the integers $a$ and $c$ above arises from the fact
that in the orbifolded theory some of the electric charges coming
from the twisted winding states are $1/N$ quantized
\cite{Vafa:1995gm,{Aspinwall:1995fw}, Sen:2005iz, Jatkar:2005bh}.

CHL orbifolds thus provide a class of reduced rank ${\cal N} =4$
compactifications that are amenable to CFT techniques. At  the same
time they have physically interesting duality symmetries. In
particular, the spectrum of dyons is required to transform correctly
under the S-duality group $\Gamma_1(N)$ which puts stringent
restrictions. The proposed partition function for the dyons in a
$\mathbb{Z}_N$ CHL orbifold that satisfies this requirement  is
given in terms of a specific Siegel modular form
\cite{Jatkar:2005bh}. Let us recall a few facts about the Siegel
forms. Let $\Omega_{ij}$, $(i, j) =1, 2$, be a $(2\times 2)$
symmetric matrix with complex entries
\begin{equation}\label{period}
   \Omega = \left(
              \begin{array}{cc}
                \rho & v \\
                v & \sigma \\
              \end{array}
            \right)\,,
\end{equation}
satisfying
\begin{equation}\label{cond1}
   \textrm{Im}(\rho) >0,\, \textrm{Im}(\sigma) >0,\, \det{(\textrm{Im}(\Omega))} >0,
\end{equation}
which parametrizes  the `Siegel upper half plane' in the space of
$(\rho, \sigma, v)$. It can be thought of as the  period matrix of a
genus two Riemann surface. For a genus-two Riemann surface, there is
a natural symplectic action of the group $Sp(2, \mathbb{Z})$ on the
period matrix.   We write an element $g$ of $Sp(2, \mathbb{Z})$ as a
$(4\times 4)$ matrix in the block form
\begin{equation}\label{sp}
   \left(
  \begin{array}{cc}
    A & B \\
    C & D \\
  \end{array}
\right),
\end{equation}
where $A, B, C, D$ are all $(2\times 2)$ matrices with integer
entries. They  satisfy
\begin{equation}\label{cond}
   AB^T=BA^T, \qquad  CD^T=DC^T, \qquad AD^T-BC^T=I\, ,
\end{equation}
so that $g^t J g =J$ where $J = \left(
                                  \begin{array}{cc}
                                    0 & -I \\
                                    I & 0 \\
                                  \end{array}
                                \right)$
is the symplectic form. The action of $g$  on the period matrix is
then  given by
\begin{equation}\label{trans}
    \Omega \to (A \Omega + B )(C\Omega + D ) ^{-1}.
\end{equation}
This connection with the genus-two Riemann surface will be important
later in $\S{\ref{Genustwo}}$, $\S{\ref{Determinants}}$, and
$\S{\ref{ZN}}$.

There is a standard embedding of $SL(2, \mathbb{Z})$ into $Sp( 2,
\mathbb{Z})$ \cite{Eichler:1985ja} used in \cite{Jatkar:2005bh}.
Using this embedding one can define a congruence subgroup $G_1(N)$
of $Sp(2, \mathbb{Z})$ that contains the $\Gamma_1(N)$ subgroup of
the $SL(2, \mathbb{Z})$. A Siegel modular form $\Phi_k(\Omega)$ of
weight $k$ and level $N$ is then defined by its transformation
property
\begin{equation}\label{phi}
    \Phi_k [(A \Omega + B )(C\Omega + D ) ^{-1}] =  \{\det{(C\Omega + D )}\}^k
    \Phi_k (\Omega), \qquad \left(
  \begin{array}{cc}
    A & B \\
    C & D \\
  \end{array}
\right) \in G_1(N).
\end{equation}
The index $k$ of the modular form is determined in terms of the
level $N$ from physics considerations,
\begin{equation}\label{relation}
    k= \frac{24}{N+1} - 2.
\end{equation}
This is necessary so that the degeneracies of the dyonic black holes
deduced from the partition function agree with the macroscopic
Bekenstein-Hawking-Wald entropy \cite{Jatkar:2005bh}. In
$\S{\ref{Determinants}}$ and $\S{\ref{ZN}}$ we give a microscopic
derivation of this relation.

The definition of $G_1(N)$ in \cite{Jatkar:2005bh} implies that
\begin{equation}\label{gone}
    C = {\bf 0}\, \textrm{mod}\, N,\quad \det{A} = 1\, \textrm{mod}\, N, \quad
    \det{D} = 1\, \textrm{mod}\, N.
\end{equation}
One can similarly define a subgroup $G_0(N)$ that contains  a
$\Gamma_0(N)$ subgroup of the $SL(2, \mathbb{Z})$. The matrices in
this group satisfy a milder condition
\begin{equation}\label{gzero}
    C = {\bf 0}\, \textrm{mod}\, N.
\end{equation}
For the purposes of physics, it is sufficient to find a modular form
that transforms as in (\ref{phi}) under $G_1(N)$ transformation
which ensures that the resulting dyonic degeneracies  transform
correctly under the $\Gamma_1(N)$ S-duality group contained in the
$G_1(N)$. In all known cases, however, the relevant Siegel modular
form $\Phi_k$ in fact transforms as in (\ref{phi}) under the larger
symmetry $G_0(N) \supset G_1(N)$. We will give an explanation of
this accidental enhanced symmetry in $\S{\ref{ZN}}$.

A dyonic state is specified by the charge vector $Q=(Q_e, Q_m)$
which transforms as a doublet of the S-duality group
$\Gamma_1(N)\subset SL(2, \mathbb{Z})$ as in (\ref{charge})and as a
vector of the T-duality group which is a subgroup of $O(22, 6;
\mathbb{Z})$. There are three T-duality invariant quadratic
combinations $Q_e^2$, $Q_m^2$, and $Q_e \cdot Q_m$ that one can
construct from these charges. The dyon degeneracies are defined not
directly in terms of $\Phi_k$ but rather in terms of $\Phi'_k$ which
is a modular form of a subgroup $ G'_0(N)$ of $Sp(2, \mathbb{Z})$
related to $G_0(N)$ by conjugation \cite{Jatkar:2005bh}. Let $g(m,
n, l)$ be the Fourier coefficients of $1/\Phi'_{k}$ defined by
\begin{equation}\label{fourier}
    \frac{C}{\Phi'_{k}(\rho, \sigma, v)} =
  \sum_{m\ge -1/N,n\ge -1, l}
   e^{2 \pi i(m \rho
     + n \sigma + l v)} g(m,n,l)\, ,
\end{equation}
where $C$ is a $N$-dependent constant
\begin{equation}\label{C}
    C = -(i \sqrt{N})^{-k-2}.
\end{equation}
The degeneracies $d(Q)$ of dyonic states of charge $Q$ are then
given by
\begin{equation}\label{degen}
    d(Q)=g\left(\frac{1}{2} Q_e^2,\, \frac{1}{2}
     Q_m^2,\, Q_e\cdot Q_m\right)\, ,
\end{equation}
The parameters $(\rho, \sigma, v)$ in the partition function can
thus be thought of as the chemical potentials for the integers
$\left(Q_e^2/2,\,  Q_m^2/2,\, Q_e\cdot Q_m\right)$ respectively. We
will see in $\S{\ref{Determinants}}$ and $\S{\ref{ZN}}$ that the
precise choice of the group $G'_0(N)$ instead of $G_0(N)$, the
relation between $k$ and $N$, as well the normalization $C$ follows
naturally from our analysis.

\section{Dyon Partition Function and  Genus-Two Riemann Surfaces \label{Genustwo}}

One puzzling feature of the proposed dyon partition function is the
appearance of the discrete group $Sp(2, \mathbb{Z})$. This group has
no direct physical significance because it cannot be viewed as a
subgroup of the U-duality group. Moreover, both $Sp(2, \mathbb{Z})$
and the period matrix $\Omega$ are objects that are naturally
related to a genus-two Riemann surface. It is equally puzzling why a
genus-two surface should play a role in the counting of dyons. An
explanation for these puzzles can be provided following the
reasoning in \cite{Gaiotto:2005hc} which we review below.

Let us first consider the toroidally compactified theory. Heterotic
on $ T^4 \times T^2$ is dual to Type-IIB on $ K3 \times T^2$. The
$SL(2, \mathbb{Z})$ symmetry which is the electric-magnetic
S-duality group in the heterotic description maps to a geometric
T-duality group in the IIB description that acts as the mapping
class group of the $ T^2$ factor. As a result, the $A$-cycle of the
$ T^2$ corresponds to electric states  and the $B$-cycle to the
magnetic states. Now, Type-IIB compactified on a small $ K3$ has an
effective 1-brane which  is a bound state of D5, NS5 branes wrapped
on K3, D3 branes wrapped on the 22 2-cycles of the $ K3$ and D1,F1
strings. A half-BPS state that is purely electric in the heterotic
description then corresponds to the 1-brane wrapping on the
$A$-cycle of the torus in the IIB description. The magnetic dual of
this state corresponds to the same 1-brane wrapping the $B$-cycle.
The 1-brane can in general carry left-moving oscillations. To count
the number of the electric states, for example, we need to count the
number of oscillating configurations of the 1-brane for a given
left-moving oscillation number. As usual, it is easier to introduce
a chemical potential conjugate to the oscillation number and compute
the partition function of this 1-brane in the canonical ensemble
with a Euclidean worldsheet and then define the microcanonical
degeneracies in terms of Fourier coefficients.

To compute the partition function, we compactify time on a Euclidean
circle with supersymmetric boundary conditions. Since Type-IIB
string on a circle is dual to M-theory on a $ T^2$,  we have a
compactification of Euclidean M-theory on $ K3 \times T^2 \times
T^2$.  Under this duality, NS5 and D5 branes of IIB map to the
M5-brane and circle-wrapped D3-branes to M2-branes. Thus the
effective 1-brane is represented by an  M5-brane wrapping a $ K3$
with  various fluxes turned on representing bound M2-branes wrapping
2-cycles of $ K3$. A K3-wrapped M5-brane is in fact dual to the
fundamental heterotic string when the $ K3$ is small by the M-theory
heterotic duality. In the low energy limit, the CFT living on the
effective 1-brane is thus the same as the one for a fundamental
heterotic string compactified on $ T^4 \times T^2$. The electric
charges are a vector $Q_e$ in the Narain lattice of signature $(22,
6)$. Level matching requires that the left-moving oscillation number
equals $Q_e^2/2$. The partition function of these half-BPS states is
then the genus-one chiral partition function of the left-moving
bosons of the heterotic string given by
\begin{equation}\label{genusone}
    Z(\tau) = \frac{1}{\eta^{24}(\tau)}
\end{equation}
where $\eta(\tau)$ is the Dedekind eta function. Given the Fourier
expansion
\begin{equation}\label{fourier2}
    Z(\tau) = \sum q^n c(n), \qquad q = \exp(2\pi i\tau),
\end{equation}
degeneracies $d(Q_e)$ are given by
\begin{equation}\label{expand}
   d(Q_e) = c(Q_e^2 /2).
\end{equation}
The modular form $\eta^{24}$ is the unique cusp form of weight $12$
of $SL(2, \mathbb{Z}) \sim Sp(1, \mathbb{Z})$ and is the inverse of
the chiral heterotic partition function on a genus-one surface. The
partition function of electric states in the CHL orbifolds can be
similarly determined
\cite{{Dabholkar:2005by},Dabholkar:2005dt,{Sen:2005pu}} and is given
by the chiral partition function of the heterotic string on a
genus-one surface with a branch cut across which the CHL bosons are
odd.

\begin{figure}
\centering
\includegraphics[width=4.9in]{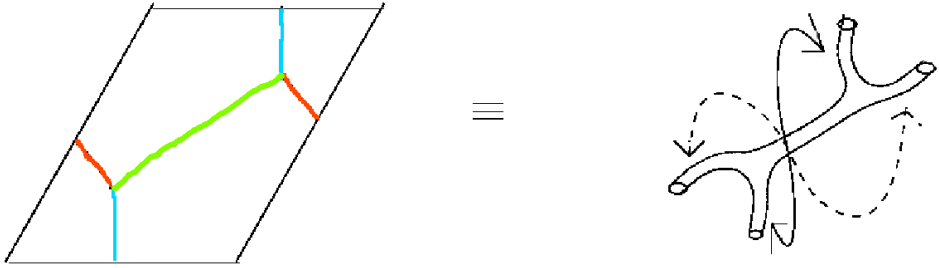}
\caption{A dyon can be represented as a string web on a torus which
in M-theory looks like a genus two Riemann surface.} \label{twoloop}
\end{figure}

For purely electric states, this long chain of dualities is of
course not necessary. We know directly that the half-BPS electric
states correspond to  the left-moving oscillations of a fundamental
heterotic string \cite{Dabholkar:1989jt, {Dabholkar:1990yf}} which
are counted by the genus-one partition function (\ref{genusone})
with appropriate orbifold projections. This chain of dualities is
however useful for generalizing the counting to quarter-BPS dyons. A
quarter-BPS dyon is a bound state of electric and magnetic charges
associated with different $U(1)$ gauge fields. In the Type-IIB
theory, this bound state is described as a string web wrapping the
torus \cite{{Aharony:1997bh}, Sen:1997xi}. The web has two vertices
and at each vertex there is three-string junction as shown in Fig.
\ref{twoloop}. At the three string junction, a 1-brane carrying
charge $Q_e$ (shown in red) combines with the other carrying charge
$Q_m$ (shown in blue) to form a 1-brane with charge $Q_e +Q_m$
(shown in green). The simplest such example to keep in mind is to
take in the IIB picture $Q_e$ to be a $ K3$-wrapped D5-brane, $Q_m$
to be the $ K3$-wrapped NS5-brane so that $Q_e +Q_m$ is a
$K3$-wrapped $(1, 1)$ 5-brane. In M-theory all 5-branes are
represented by a single M5-brane wrapping $K3$. The topology of the
quarter-BPS web of the effective 1-brane wrapping a 2-torus with two
vertices is that of a genus two Riemann surface after adding the
Euclidean time circle. In the dual M-theory, the 1-brane wrapped on
the torus is dual to an M5 brane wrapping $K3$ and a holomorphic
curve in the $T^4$ and the string junctions are smooth in this
description. Thus the low-energy description of a quarter-BPS brane
is a Euclidean M5-brane wrapping $ K3 \times \Sigma_2$ embedded in $
K3 \times T^4$. This embedding is achieved by the natural
holomorphic embedding of a genus two Riemann surface into a $ T^4$
given by the Abel map which maps a complex curve into its Jacobian.
The dyonic partition function is then given by the sum over all
`left-moving' fluctuations of this worldvolume. Since K3-wrapped
M5-brane is the heterotic string we are thus led to computing the
genus-two partition function of the left-moving fluctuations of the
heterotic string.\footnote{On higher genus Euclidean surfaces there
is no isometry to define left and right. We define the `left-moving
partition function'  as usual by holomorphic factorization taking
the holomorphic square root of the genus two partition function of
the bosonic string.}

Consistent with this picture it is known that the Igusa cusp form
$\Phi_{10}$ that appears in the dyon partition function in this case
with $N=1$ is precisely the left-moving genus-two partition function
of the toroidally compactified heterotic string \cite{Moore:1986rh,
Belavin:1986tv}. Note that at genus two, the ghost determinants and
the light-cone directions do not quite cancel out. Thus, the fact
that one obtains a Siegel form of weight $10$ depends nontrivially
on the ghosts and the light-cone directions and does not follow
merely from the $24$ transverse directions.\footnote{Note that this
reasoning gives us a description of quarter BPS dyons in terms of
fluctuations of string webs and heterotic world sheets in  static
gauge. The computation of fluctuation determinants is most
conveniently performed in covariant gauge. We are implicitly
assuming the equivalence between covariant and static gauge.}

The same conclusion can be reached by a slightly different reasoning
in a single duality step within the heterotic description. For
computing a partition function, we take time to be a Euclidean
circle with supersymmetric boundary conditions. Both the fundamental
string and the NS5-brane component extend along the euclidean time.
We now have effectively, heterotic string on $ T^7$ which has a
larger U-duality group $O(8, 24; \mathbb{Z})$ that combines together
the T-duality group $O(7, 23; \mathbb{Z})$ and the $SL(2,
\mathbb{Z})$ S-duality group \cite{Sen:1994wr}. The chain of
dualities described above can be condensed into a single U-duality
transformation that sends the Euclidean NS5-branes wrapped along
time to Euclidean fundamental strings that do not wrap the time
circle, while at the same time leaving the fundamental strings
wrapped along time unchanged. As a result, the partition function of
quarter-BPS dyons is mapped to the partition function of a heterotic
string worldsheet. The worldsheet has a component that wraps around
time and one spatial direction, and another that wraps around the
other two spatial directions, so that it wraps effectively a genus
two curve.

Using this picture it is also easy to see why $\rho$, $\sigma$ and
$v$ are the chemical potentials associated with $Q_e^2/2$, $Q_m^2/2$
, and $Q_e \cdot Q_m$ respectively. The genus-two momentum sum takes
the form \eqref{classsum} with $(p^i_L, p^i_R)$ taking values in
$\Gamma^{22, 6}$ Narain lattice in the new heterotic frame. The
momenta $(p^1_L, p^1_R)$ are identified with $Q_e$ of the original
heterotic frame whereas the momenta $(p^2_L, p^2_R)$ are identified
with $Q_m$. Integrating along $0\leq \textrm{Re}(\rho) \leq 1$,
$0\leq \textrm{Re}(\sigma) \leq 1$, $0\leq \textrm{Re}(v) \leq 1$ as
in \cite{Dijkgraaf:1996it} means $\Omega$ is real along the
integration contour. {}From these facts it follows from the form of
\eqref{classsum} that $\rho$, $\sigma$ and $v$ are the chemical
potentials associated with $Q_e^2/2$, $Q_m^2/2$ , and $Q_e \cdot
Q_m$ respectively.

So far we have been discussing the toroidally compactified case with
$N=1$. For CHL orbifolds for other values of $N$, the genus-two
worldsheet would have a branch cut of order $N$ along one of the
cycles. This can be seen most easily in the string web picture. The
CHL orbifold action combines an order $N$ shift $T$ along one of the
circles with an order $N$ left-moving twist $\xi$ of the internal
CFT. This implies that to construct a state wrapping a compact $
T^2$ from the string web, two ends of the web in the fundamental
cell of the $ T^2$ along one of the cycles (shown in blue for
example in Fig. \ref{twoloop}) are joined after an order $N$ twist
$\xi$. Using M-theory lift and heterotic dual as above, the
resulting genus-two worldsheet then has a branch cut across which
some of the left-moving fields undergo a $\mathbb{Z}_N$ twist. This
will be explained more concretely for the $\mathbb{Z}_2$ case in the
next section. Computation of the genus-two partition function will
then reveal the correct Siegel modular form consistent with this
picture.

\section{Computation of the Twisted  Chiral Partition Function \label{Determinants}}

We would like to  confirm the  picture in the previous section with
a  computation of the genus-two partition function for CHL orbifolds
for other values of $N >1$. We focus in this section on the
$\mathbb{Z}_2$ CHL model where our computations can be carried
through completely and where closed form expressions for $\Phi_6$
are available in terms of $\Phi_{10}$ and theta functions. This will
allow for an explicit comparison of our computation with the
proposed partition function. These considerations naturally
generalize to other values of $N$.

To construct the $\mathbb{Z}_2$ CHL orbifold, one starts with a
toroidally compactified $E_8 \times E_8$ string which admits a
$\mathbb{Z}_2$ left-moving symmetry generated by the element $\xi$
that flips the two $E_8$ factors. In the bosonic representation of
the $E_8 \times E_8$ current algebra, the first and the second $E_8$
factors can be represented by left-moving bosons $X^I$ and $Y^I$
respectively each living on the $E_8$ root lattice with $I= 1,
\ldots 8$. The combination $(X^I + Y^I)/\sqrt{2}$ is then even under
$\xi$ and $(X^I - Y^I)/\sqrt{2}$ is odd. The CHL orbifolding action
then combines this with a half-shift $T$ along $ S^1$. We are
interested in the genus-two chiral partition function of the
left-moving bosons. The genus two worldsheet for the twisted sectors
has a branch cut across which the fields $(X^I - Y^I)/\sqrt{2}$ flip
sign.

To evaluate the  partition function of bosons on a genus $g$ Riemann
surface with branch cuts, we closely follow the discussion in
\cite{Dijkgraaf:1987vp}. The partition function in this case can be
written as a product of a classical piece and a quantum piece. The
classical piece comes from a sum over instanton sectors of classical
maps that have nontrivial windings around the target space and gives
rise to a theta function over the Narain lattice. The quantum piece,
which is usually the more difficult piece to evaluate,  is
determined in terms of  the fluctuation determinants of the scalars
as well as the ghosts. The bosonic determinants in general have
complicated expressions and it is not easy to extract them in a
useful form that can be compared with the Siegel modular forms. To
circumvent this difficulty, we express the twisted partition
function in terms of the untwisted partition function and theta
functions.  Since the total untwisted partition function is given as
the inverse of the Igusa cusp form $\Phi_{10}$, we can then extract
a closed form expression for the total twisted partition function.
Our strategy will be to treat the quantum and classical pieces
separately. We first evaluate the quantum piece and will later treat
the classical piece that gives the lattice sum over charges. The two
pieces individually have a dependence on Prym periods which cancels
out from the final answer.

The quantum piece of the twisted theory can be obtained from the
quantum piece of the untwisted theory by replacing eight untwisted
bosonic determinants by eight twisted determinants,
\begin{equation}\label{twistun}
    Z^{qu}_{twisted} = Z^{qu}_{untwisted}\, (\frac{\Delta_{u}}
    {\Delta_{t}})^8.
\end{equation}
Here $\Delta_{u}$ and $\Delta_{t}$ are the determinants over nonzero
modes  of the $\bar \partial$ operator on a genus two surface with
untwisted and twisted boundary conditions across the branch cut
respectively.

To evaluate the ratio of determinants, it is sufficient to consider
a single scalar field $\varphi$ compactified on a circle of radius
$R$ on a genus two Riemann surface $\Sigma$. We choose a canonical
homology basis  $(A_i, B_i)$, $i = 1, 2$ for the surface $\Sigma$,
normalized with respect to the intersection product
\begin{equation}\label{inter}
   \# (A_i, A_j) = \#(B_i, B_j) =0; \quad \#(A_i, B_j) = \#(B_i, A_j)
   = \delta_{ij}.
\end{equation}
Note that linear relabeling of the homology cycles that leaves the
intersection product invariant generates  precisely the  $Sp(2,
\mathbb{Z})$ group of matrices defined in (\ref{sp}).

Let us first consider the unorbifolded theory of a scalar $\varphi$
on a circle, so that the Riemann surface $\Sigma$ has no branch
cuts. The partition function is then given by the product of the
classical and the quantum piece
\begin{equation}\label{Z}
    Z_{\textrm{circle}}(R) = Z_{\textrm{circle}}^{cl}(R)  Z_{\textrm{circle}}^{qu}\, .
\end{equation}
The entire $R$ dependence is in the classical piece. Explicit form
of the quantum part $Z^{qu}$ will not be needed for our purpose. The
classical part is given by the partition sum over classical
solutions $\partial \bar
\partial \varphi^{cl} =0$ with winding numbers around
various cycles. The nontrivial classical solutions can be written in
terms of the integrals of the holomorphic one-forms $\omega_i =
\omega_i(z) dz$ and their complex conjugates $ \bar \omega_i
=\bar\omega_i(\bar z)d\bar z$, $i = 1,2$. The holomorphic one-form
are normalized with respect to the $A_i$ cycles and define the
period matrix $\Omega$ by
\begin{equation}\label{periodmat}
    \oint_{A_i} \omega_j = \delta_{ij} \qquad \oint_{B_i} \omega_j =
    \Omega_{ij}.
\end{equation}
The winding numbers of $(l_i, m_i)$ the classical solution are
defined by the periods
\begin{equation}\label{cycles}
    \oint_{A_i} d\varphi^{cl} = 2\pi R m_i,
    \quad \oint_{B_i} d\varphi^{cl} = 2\pi R l_i.
\end{equation}
A solution with given winding numbers can be written as
\begin{equation}\label{classi2}
   \varphi^{cl}(z) =-\frac{1}{2} \pi i R(l - \bar \Omega m) \cdot
   (\textrm{Im}
   \Omega)^{-1} \cdot \int_{z_0}^z \omega  + \textrm{c}. \textrm{c},
\end{equation}
with classical action
\begin{equation}\label{classiact}
    S[\varphi^{cl}] = \frac{1}{2} \pi R^2 (m - \bar \Omega l)\cdot
    (\textrm{Im} \Omega)^{-1} \cdot (m - \Omega l).
\end{equation}
Partition sum over these winding sectors then gives
\begin{equation}\label{classsum}
Z^{cl}(R) = \sum_{(p_L^i,p_R^i)} \exp[\pi i (p_L \cdot \Omega \cdot
p_L- p_R \cdot \bar \Omega  \cdot p_R)],
\end{equation}
after Poisson resummation.  The  lattice sum can be interpreted as
the sum  over the left and right-moving momenta $\{p_L^i, p_R^i\}$
flowing through each $B_i$ cycle. For each $i$, the momenta take
values in  self-dual Lorentzian Narain lattice $\Gamma^{1,1}$
spanned by
\begin{equation}\label{momenta}
    p^i_{R, L} =(\frac{n^i}{R} \pm \frac{w^i
    R}{2}).
\end{equation}
We use the standard CFT conventions throughout so $\alpha' =2$, the
self-dual radius is $R = \sqrt{2}$, and the free field OPE are
normalized $\phi(z) \phi(w) \sim -\ln(z-w)$.

In the orbifolded theory, the  partition function is obtained by a
functional integral over fields $\varphi$ which can be double-valued
on $\Sigma$.  These field configurations can be even or odd going
around the four cycles $(A_i, B_i)$ and fall into $2^4$ distinct
topological sectors corresponding to elements of $H_1(\Sigma,
\mathbb{Z}_2)$ labeled by the half-characteristics
\begin{equation}\label{char}
    \left[\begin{array}{c}
            \epsilon \\
            \delta
          \end{array}\right] = \left[\begin{array}{c c}
            \epsilon_1 &  \epsilon_2\\
            \delta_1 &  \delta_2
          \end{array}\right], \qquad (\epsilon_i, \delta_i = 0,
          \frac{1}{2}\,\,  \textrm{for}\,\, i= 1, 2).
\end{equation}
The orbifold partition function is then a sum over all sectors of
the genus two surface,
\begin{equation}\label{orbpart}
    Z_{\textrm{orbifold}}(R) = \frac{1}{2^2} \sum_{\epsilon, \delta}
    Z_{\epsilon, \delta}(R),
\end{equation}
where $Z_{\epsilon, \delta}(R)$ is a partition sum over field
configuration that are double valued across a branch cut along the
cycle $\sum_i 2 (\delta_i A_i + \epsilon_i B_i)$. As before, in each
sector the partition sum is a product of a quantum piece and a
classical piece.
\begin{equation}\label{twistedpart}
  Z_{\epsilon, \delta}(R) = Z^{qu}_{\epsilon, \delta} Z^{cl}_{\epsilon, \delta}(R)
\end{equation}
The untwisted partition function $Z_{0,0}(R)$ is the partition
function $Z_{\textrm{circle}}(R)$ of the circle theory that we have
already evaluated above. For a nonzero  characteristics, we can
choose a new homology basis so that
\begin{equation}\label{char2}
    \left[\begin{array}{c}
            \epsilon \\
            \delta
          \end{array}\right] = \left[\begin{array}{c c}
            0 &  0\\
            0 &  \frac{1}{2}
          \end{array}\right].
\end{equation}
which corresponds to choosing the branch cut around the new $A_2$
cycle. This branch cut defines a double cover ${\bf \pi}: \hat
\Sigma \rightarrow \Sigma$ obtained by cutting the Riemann surface
along the cycle $A_2$ and then pasting it with an identical copy
which corresponds to the second Riemann sheet. The covering Riemann
surface $\hat \Sigma$ is of genus three which is uniquely determined
by the choice of the branch cut given a $\Sigma$. The double cover
$\hat \Sigma$ admits a conformal involution ${\emph{i}}: \hat \Sigma
\rightarrow \hat \Sigma$, satisfying $\pi\circ {\emph{i}} = \pi$,
which  basically interchanges the two Riemann sheets.

\begin{figure}
\centering
\includegraphics[width=3.9in,angle=0]{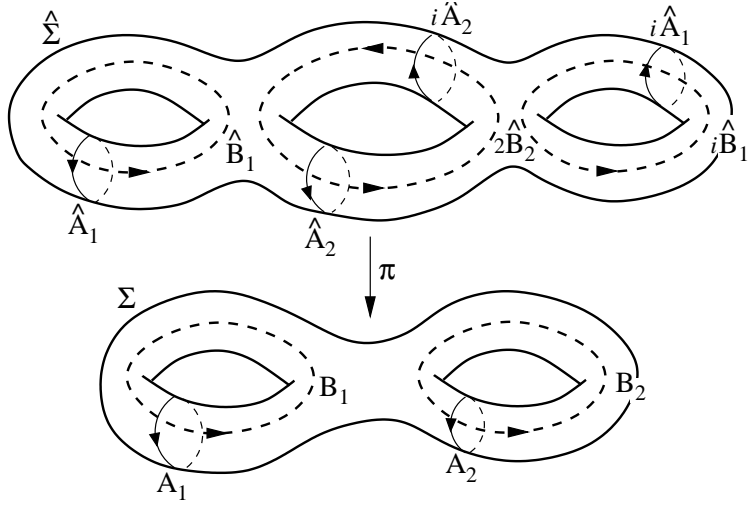}
\caption{A double cover of a genus-two Riemann surface is a
genus-three surface that admits an involution {\emph{i}}.}
\label{doublecover}
\end{figure}

A convenient choice for the homology basis on $\hat\Sigma$ is the
one that projects on the homology basis of $\Sigma$. So it is given
by $A_1, B_1;  A_2, 2 B_2; {\emph{i}}(A_1), \emph{i}( B_1)$. Given
this double cover, one can define the Prym differential $\nu
=\nu(\hat z) d\hat z$ that is odd under the defining involution
$\nu(\emph{i}(\hat z))= - \nu(\hat z)$. They are normalized with
respect to the A-cycles and defines the Prym period $\tilde \tau$
by,
\begin{equation}\label{prym}
    \oint_{\hat A_1} \nu =  -\oint_{\emph{i}(\hat A_1)} \nu = 1; \qquad
    \oint_{\hat B_1} \nu = -\oint_{\emph{i}(\hat B_1)} \nu = \tilde \tau
\end{equation}
The Prym differential has no periods around $\hat A_2$ and $ 2\hat
B_2$.  It is clear from the definition that the projection of the
Prym differential on $\Sigma$ is a double valued holomorphic
one-form which is antiperiodic around $B_2$.  More generally, for a
genus $g$ Riemann surface $\Sigma$, the double cover $\hat \Sigma$
is a Riemann surface of genus $(2g -1)$ and there are $(g-1)$ Prym
differentials. Basically, the $(2g-1)$ holomorphic forms on the
covering surface $\hat \Sigma$ can be split into even and odd under
the automorphism. The $g$ even ones project down to the holomorphic
differentials on $\Sigma$, the $(g-1)$ odd ones are the Prym
differentials. The Prym periods in this case are a $(g-1) \times
(g-1)$ matrix. For further details in this context see
\cite{Dijkgraaf:1987vp}.

The twisted sector partition function on $\Sigma$ with the twist
characteristic $(\epsilon, \delta )$ is then given by a functional
integral over field configurations $\hat \varphi$ on the double
cover $\hat \Sigma$ that are odd under the involution, $\hat
\varphi( \emph{i}(\hat z)) = - \hat\varphi(\hat z)$ mod $2\pi R$.
Because of the doubled area of $\hat \Sigma$, the action is halved
compared to the standard normalization. The  instanton
configurations on the worldsheet $\hat \Sigma$ that contribute to
the classical piece of the partition function are analogous to the
untwisted ones and have the form
\begin{equation}\label{classi}
   \varphi^{cl}(z) =-\frac{1}{2} \pi i R(l - \bar{\tilde\tau} m)
   (\textrm{Im}
   \tilde\tau)^{-1} \int_{{\emph i}(\hat z)}^{\hat z} \nu  + \textrm{c}.
   \textrm{c}.
\end{equation}
The twisted partition function in (\ref{twistedpart}) then takes the
form
\begin{equation}\label{prympartition}
    Z_{\epsilon, \delta}(R) = Z^{qu}_{\epsilon, \delta}\sum_{(p_L,p_R)}
    \exp[\pi i (p_L \cdot {\tilde\tau_{\epsilon, \delta}} \cdot
p_L- p_R \cdot \bar{\tilde\tau}_{\epsilon, \delta}  \cdot p_R)],
\end{equation}
where the sum is over a single copy of the Narain lattice
$\Gamma^{1,1}$ (\ref{momenta}) and ${\tilde\tau}_{\epsilon, \delta}$
is the Prym period corresponding to the characteristic $(\epsilon,
\delta)$.

Computing the quantum piece $Z^{qu}_{\epsilon,\delta}$ is much
harder but it will suffice for our purposes to know the ratio of the
quantum pieces for the twisted and untwisted bosons. This can be
computed using the following trick that exploits the extra $SU(2)$
symmetry available for a free boson at the self-dual radius. The
three $SU(2)$ currents $(J_x, J_y, J_z)$  at $R = \sqrt{2}$ are
given by $(\cos \sqrt{2}X, \sin \sqrt{2}X, \frac{1}{\sqrt{2}}
\partial X )$.  The $\mathbb{Z}_2$ orbifold action $X \to -
X$ is a Weyl reflection that acts as a rotation through $\pi$ around
the $x$ axis. This flips the sign of both $J_y$ and $J_z$ but leaves
$J_x$ unchanged. This is clearly equivalent by $SU(2)$ conjugation
to a rotation around the $z$ axis, which flips the sign of both
$J_x$ and $J_y$ but leaves $J_z$ unchanged. Such a  flip is achieved
by a half shift on the circle $X \to X + \pi \sqrt{2}$. An orbifold
by a half shift results in  a circle conformal field theory at half
the radius. Hence we have the equality between the orbifold CFT and
the circle CFT as follows
\begin{equation}\label{orb}
    Z_{\textrm{orbifold}}(R = \sqrt{2}) = Z_{\textrm{circle}}
    (R = \frac{1}{\sqrt{2}}).
\end{equation}

To use this equality effectively, it is convenient to write the
non-holomorphic lattice sums in (\ref{classsum}) and
(\ref{prympartition})  as  sums over holomorphically factorized
chiral blocks. For this purpose, recall that a genus $g$ theta
constants with characteristics  are defined by
\begin{equation}\label{theta}
    \vartheta\left[
             \begin{array}{cc}
                \alpha\\
                \beta\\
             \end{array}
           \right](\Omega) =
           \sum_{n\in \mathbb{Z}^g} \exp[i\pi(n+
           \alpha)\cdot \Omega \cdot (n + \alpha) + 2\pi i (n +
           \alpha) \cdot  \beta],
\end{equation}
where the characteristics $\alpha$, $\beta$ are in general
$g$-dimensional vectors and  $\Omega$ is a $(g \times g)$ period
matrix. When the radius $R^2/2$ is a rational number $p/q$, the
momentum lattices appearing in (\ref{classsum}) and
(\ref{prympartition}) can be built up from a finite number of square
sublattices. As a result, one can express the classical sum in
(\ref{classsum}) for example as
\begin{eqnarray}
  Z^{cl}(R) &=& \sum_{\alpha, \beta, \gamma} \vartheta \left[\begin{array}{c}
                                                             \frac{1}{2} \alpha + \frac{1}{2}\beta +\gamma \\
                                                               0
                                                             \end{array}\right](2pq\Omega)
                                                             \overline{\vartheta \left[\begin{array}{c}
                                                               \frac{1}{2} \alpha - \frac{1}{2}\beta +\gamma \\
                                                               0
                                                             \end{array}\right](2pq\Omega)}
   \\
  &=& 2^{-2}\sum_{\alpha, \beta, \gamma} e^{4\pi i \beta\cdot\gamma} \vartheta \left[\begin{array}{c}
                                                               \alpha + \beta \\
                                                               \gamma 
                                                             \end{array}\right](\frac{1}{2}pq\Omega)
                                                             \overline{\vartheta \left[\begin{array}{c}
                                                                \alpha - \beta \\
                                                               \gamma
                                                             \end{array}\right](\frac{1}{2}pq\Omega)},
\end{eqnarray}
where the summation is over $\alpha\in
\left(\frac{1}{p}\mathbb{Z}_p\right)^g$, $\beta\in
\left(\frac{1}{q}\mathbb{Z}_q\right)^g$, $\gamma\in (\frac{1}{2}(
\mathbb{Z}_2)^g$. Applying this formula to  our cases of interest,
we have
\begin{equation}\label{orbifoldsum}
    Z^{cl}_{\epsilon, \delta}(\sqrt{2}) =
    \sum_{\alpha}\Bigg\vert\vartheta\left[\begin{array}{c}
                                             \alpha \\
                                             0
    \end{array}\right](2{\tilde\tau}_{\epsilon, \delta})\Bigg\vert^2
\end{equation}
for the orbifold classical sum in (\ref{orbpart}) which depends on a
genus-one Prym period ${\tilde \tau}_{\epsilon, \delta}$ and
\begin{equation}\label{circlesum}
    {Z^{cl}}_{\textrm{circle}}(\frac{1}{\sqrt{2}}) = \frac{1}{2^2}
    \sum_{\alpha, \beta, \gamma}
                            \Bigg\vert\vartheta\left[\begin{array}{c}
                                             \alpha + \frac{1}{2} \gamma\\
                                             \beta
    \end{array}\right](2\Omega)\Bigg\vert ^2
\end{equation}
for the circle classical sum which depends on the genus-two period
matrix $\Omega$. The equality (\ref{orb}) then reads
\begin{equation}\label{equality}
    \frac{1}{2^2} \sum_{\epsilon, \delta, \gamma}
    Z^{qu}_{\epsilon, \delta}\Bigg\vert\vartheta\left[\begin{array}{c}
                                                   \gamma \\
                                                   0
                                                 \end{array}\right]
    (2\tilde\tau_{\epsilon, \delta}) \Bigg\vert ^2 =
    \frac{1}{2^2} \sum_{\epsilon, \delta, \gamma}
    Z^{qu}_{0, 0}\Bigg\vert\vartheta\left[\begin{array}{c}
                                                  \frac{1}{2}\epsilon +
                                                  \gamma \\
                                                   \delta
                                                 \end{array}\right]
    (2\Omega) \Bigg\vert ^2 \, .
\end{equation}
A term by term comparison of the chiral blocks following the
arguments in \cite{Dijkgraaf:1987vp} then leads to the desired
expression for the twisted determinant in terms of the untwisted
determinant
\begin{equation}\label{partitioned}
    Z^{qu}_{\epsilon, \delta} = \Bigg\vert c\left[
                                    \begin{array}{c}
                                      \epsilon \\
                                      \delta \\
                                    \end{array}
                                  \right]\Bigg\vert ^{-2} Z^{qu}_0,
\end{equation}
with
\begin{equation}\label{c}
    c\left[
       \begin{array}{cc}
         0 & 0 \\
         0 & \frac{1}{2} \\
       \end{array}
     \right]= \frac{\vartheta\left[
                            \begin{array}{c}
                              \gamma \\
                              0 \\
                            \end{array}
                          \right](2 {\tilde \tau})}{\vartheta \left[
                                               \begin{array}{cc}
                                                 \gamma & 0 \\
                                                 0 & \frac{1}{2} \\
                                               \end{array}
                                             \right](2\Omega)}.
\end{equation}
For comparison with literature we use the notation $\vartheta_{abcd}
= {\vartheta \left[
                                               \begin{array}{cc}
                                                 \frac{a}{2} & \frac{b}{2} \\
                                                 \frac{c}{2} & \frac{d}{2} \\
                                               \end{array}
                                             \right]}$ and
$\vartheta_{ab} = {\vartheta\left[
                            \begin{array}{c}
                              \frac{a}{2} \\
                              \frac{b}{2} \\
                            \end{array}
                          \right]}$.
There are useful doubling identities for theta functions,
\begin{equation}\label{doubling}
  \vartheta^2_{10}(\tilde \tau)
  = 2 \vartheta_{00}(2 \tilde \tau) \vartheta_{10}(2 \tilde \tau),
  \quad \vartheta^2_{00}(\tilde \tau)
  = \vartheta_{00}(2 \tilde \tau)^2+ \vartheta_{10}(2 \tilde \tau)^2,
  \quad \vartheta^2_{01}(\tilde \tau)
  = \vartheta^2_{00}(2 \tilde \tau) - \vartheta^2_{10}(2 \tilde
  \tau),
\end{equation}
and similarly for the genus-two theta functions. Squaring or
multiplying the two expressions for $c$ in (\ref{c}) and rearranging
them using doubling identities,  we get other useful expressions
\begin{equation}\label{csquare}
    c^2\left[
       \begin{array}{cc}
         0 & 0 \\
         0 & \frac{1}{2} \\
       \end{array}
     \right] = \frac{\vartheta^2\left[
                            \begin{array}{c}
                              \alpha \\
                              \beta \\
                            \end{array}
                          \right]({\tilde \tau})}{\vartheta \left[
                                               \begin{array}{cc}
                                                 \alpha & 0 \\
                                                 \beta & 0 \\
                                               \end{array}
                                             \right](\Omega)
                                             \vartheta \left[
                                               \begin{array}{cc}
                                                 \alpha & 0 \\
                                                 \beta & \frac{1}{2} \\
                                               \end{array}
                                             \right](\Omega)},
\end{equation}
that are available in \cite{Dijkgraaf:1987vp}. Note that the left
hand side is independent of the characteristics $\alpha$ and
$\beta$. Equality of the right hand side for different choices of
$\alpha$ and $\beta$ is  the statement of the Schottky relations
\cite{Dijkgraaf:1987vp}. Using different values of $\alpha$ and
$\beta$ we obtain
\begin{equation}
c^2_{0001}=\frac{\vartheta_{00}^2(\tilde
\tau)}{\vartheta_{0000}(\Omega)\vartheta_{0001}(\Omega)}=
\frac{\vartheta_{01}^2(\tilde
\tau)}{\vartheta_{0010}(\Omega)\vartheta_{0011}(\Omega)}=
\frac{\vartheta_{10}^2(\tilde
\tau)}{\vartheta_{1000}(\Omega)\vartheta_{1001}(\Omega)} \, .
\end{equation}
Using these three equations we can write
\begin{equation}\label{c1}
   \frac{1}{c^{8}_{0001}} = \frac{\vartheta_{0000}^2(\Omega)
\vartheta_{0001}^2(\Omega)
\vartheta_{0010}^2(\Omega)\vartheta_{0011}^2(\Omega)}
{\vartheta_{00}^4(\tilde \tau)\vartheta_{01}^4(\tilde \tau)}\, ,
\end{equation}

\begin{equation}\label{c2}
   \frac{1}{c^{8}_{0001}} = \frac{\vartheta_{0000}^2(\Omega)
\vartheta_{0001}^2(\Omega)
\vartheta_{1000}^2(\Omega)\vartheta_{1001}^2(\Omega)}
{\vartheta_{00}^4(\tilde \tau)\vartheta_{10}^4(\tilde \tau)}\, ,
\end{equation}

\begin{equation}\label{c3}
    \frac{1}{c^{8}_{0001}}  = \frac{\vartheta_{1000}^2(\Omega)\vartheta_{1001}^2(\Omega)
\vartheta_{0010}^2(\Omega)\vartheta_{0011}^2(\Omega)}{\vartheta_{10}^4(\tilde
\tau)\vartheta_{01}^4(\tilde \tau)}\, .
\end{equation}
These expressions are now in a convenient form to compare with the
known expressions for $\Phi_{6}$ \cite{Ibukiyama:1991ibu3,
AI:2005ai1}

\begin{equation}\label{ibuk}
\frac{1}{\Phi_{6}(\Omega)} =
\frac{\vartheta_{0000}^2(\Omega)\vartheta_{0001}^2(\Omega)
\vartheta_{0010}^2(\Omega)\vartheta_{0011}^2(\Omega)}{\Phi_{10}(\Omega)}\,
,
\end{equation}

\begin{equation}
\frac{1}{\Phi'_{6}(\Omega)} =
\frac{\vartheta_{0000}^2(\Omega)\vartheta_{0001}^2(\Omega)
\vartheta_{1000}^2(\Omega)\vartheta_{1001}^2(\Omega)}{\Phi_{10}(\Omega)}\,
,
\end{equation}

\begin{equation}
\frac{1}{\Phi''_{6}(\Omega)} =
\frac{\vartheta_{1000}^2(\Omega)\vartheta_{1001}^2(\Omega)
\vartheta_{0010}^2(\Omega)\vartheta_{0011}^2(\Omega)}{\Phi_{10}(\Omega)}\,
.
\end{equation}
Here $\Phi_{6},\Phi'_{6},\Phi''_{6}$ are various images of
$\Phi_{6}$ under an $SL(2,\mathbb{Z})$ subgroup of the $Sp(2,
\mathbb{Z})$ modular transformations. There are only three images
because $\Phi_{6}$ is invariant under the subgroup $\Gamma_0(2)$ of
index $3$ in $SL(2,\mathbb{Z})$.

To compare with (\ref{ibuk}), for example, one can multiply the expression (\ref{c1}) with the inverse power of $\Phi_{10}(\Omega)$.
We therefore conclude that the quantum piece of the twisted chiral
partition function of the left-moving heterotic string with the
$\mathbb{Z}_2$ branch cut along the $A_2$ cycle is given by
\begin{equation} \label{expression}
\frac{\Delta^8_{0000}(\Omega)}{\Delta^8_{0001}(\Omega)}
\frac{1}{\Phi_{10}(\Omega)} =
\frac{1}{c^8_{0001}}\frac{1}{\Phi_{10}(\Omega)} =
\frac{1}{\vartheta_{00}^4(\tilde \tau)\vartheta_{01}^4({\tilde\tau})
\Phi_6(\Omega)},
\end{equation}
with similar expressions in terms of $\Phi'_6$ and $\Phi''_6$. This
is promising but still not quite right because there  is an unwanted
factor in the denominator involving theta functions over the Prym
periods. As we will see below, these factors are eliminated once the
classical contribution from the momentum sum is properly taken into
account.

In the orbifolded theory, there are no gauge fields that couple to
the odd $E_8$ charges. Nevertheless, states with these charges still
run across the $B_1$ cycle of the genus two surface in
Fig.\ref{doublecover} that has no branch cut. Although we want to
fix the momenta running along the two handles of the Riemann surface
in terms of the electric and magnetic charges of the dyon, we still
have to sum over the appropriate lattice of odd charges. The sum
over the odd charges will give a theta function of the Prym period.
We will show that it exactly cancels the theta function appearing in
the denominator in (\ref{expression}).

The numerator for genus-two twisted partition function for the
$\mathbb{Z}_2$ orbifold of the $E_8 \times E_8$ lattice is a
combination of expressions like (\ref{classsum}) for the even
momentum lattice and (\ref{prympartition}) for the odd momentum
lattice. The lattice sum is best understood in terms of the covering
genus-three surface $\hat \Sigma$. In the basis of cycles chosen
earlier, it follows from the definition that the genus-three period
matrix can be expressed in terms of genus-two quantities as
\begin{equation}\label{sigmahat}
    \hat \Omega = \left(
       \begin{array}{ccc}
         \frac{1}{2}(\rho + \tilde \tau) & v &
         \frac{1}{2}(\rho - \tilde \tau) \\
         v & 2\sigma & v  \\
         \frac{1}{2}(\rho - \tilde \tau) & v &
         \frac{1}{2}(\rho + \tilde \tau) \\
       \end{array}
     \right).
\end{equation}
Upon projection to $\Sigma$, the period matrix gives rise to the
genus-two period matrix and the Prym period. The genus-three
momentum sum over the $E_8 \times E_8$ lattice $\Lambda$ is
\begin{equation}\label{momsumhat}
    \sum_{{\bf p} \in \Lambda^3} \exp{[\pi i ({\bf p} \cdot \hat \Omega \cdot {\bf
    p})]}.
\end{equation}
Given the form (\ref{sigmahat}) of the period matrix, this
expression can be reorganized in terms of genus-two objects as
explained in appendix A  to obtain ,
\begin{eqnarray}
\nonumber \Theta_{E8}(2\rho, 2\sigma,
2v)\vartheta^4_{00}({\tilde\tau}) \vartheta^4_{01}({\tilde\tau}) +
\frac{1}{16}\Theta_{E8}(\frac{\rho}{2}, 2\sigma,
v)\vartheta^4_{00}({\tilde\tau}) \vartheta^4_{10}({\tilde\tau}) -
\frac{1}{16}\Theta_{E8}(\frac{\rho+1}{2}, 2\sigma,
v)\vartheta^4_{01}({\tilde\tau}) \vartheta^4_{10}({\tilde\tau}).
\end{eqnarray}
Combining this classical piece with the quantum twisted determinants
that we have calculated earlier, we see that the dependence on Prym
periods cancels completely. The full twisted partition function is
then
\begin{equation}\label{twistefinal}
Z_{twisted}= \frac{\Theta_{E8}(2\rho, 2 \sigma,  2 v
)}{\Phi_6(\Omega)} + \frac{1}{16}\frac{\Theta_{E8}(\frac{\rho}{2}, 2
\sigma, v)}{\Phi'_6(\Omega)} - \frac{1}{16}\frac{
\Theta_{E8}(\frac{\rho+1}{2}, 2 \sigma, v)}{\Phi''_6(\Omega)}.
\end{equation}
Here we have suppressed the uninteresting momentum sum coming from
the toroidal compactification.

Recall that the  proposal of \cite{Jatkar:2005bh} gives the dyon
partition function for a specific class of states, which carry
half-integral units of winding along the CHL circle. Comparison with
the genus one partition function for electric states computed for
example in \cite{Dabholkar:2005dt} shows that the states that carry
half a unit of winding on the CHL circle are associated with the
even $E_8$ charges contained in $\Theta_{E8}(\rho/2,  2 \rho, v )$.
Hence we reproduce the result that the corresponding degeneracies
are the Fourier coefficients of $-1/ 16 {\Phi'_6(\Omega)}$.
Comparing the expression for $C$ in \eqref{C} for $N=1$ and $N=2$,
we see that we also correctly reproduce the normalization relative
to the $N=1$ case.

This concludes the main part of the derivation. The Igusa cusp form
$\Phi_{10}$ is proportional to the product of all ten  genus-two
theta functions with even spin structure,
\begin{equation}\label{igusa}
    \Phi_{10}(\Omega) = 2^{-12} \prod_{{\alpha, \beta}\atop
   4\alpha\cdot\beta = even}  \vartheta^2\left[
             \begin{array}{cc}
                \alpha\\
                \beta\\
             \end{array}
           \right](\Omega).
\end{equation}
This allows us to express $\Phi_6$ using \eqref{ibuk} as a product
of six particular even theta functions
\begin{equation}\label{ibukiyama}
    \Phi_6=  2^{-12} (\vartheta_{0100} \vartheta_{0110}
    \vartheta_{1000} \vartheta_{1001}\vartheta_{1100} \vartheta_{1111})^2 ,
\end{equation}
Our results can thus be viewed as a CFT derivation of the level 2
Siegel modular form $\Phi_6$ in terms of theta functions that was
obtained by Ibukiyama from a  very different starting point
\cite{Ibukiyama:1991ibu3,{AI:2005ai1}}.

A few comments are in order. First, let us make the modular
properties of $\Phi_6(\Omega)$ under $G_0(2)$ more manifest.  The
theta function in the numerator, $\Theta_{E8}(2 \Omega)$ is clearly
modular under $G_0(2)$. Note that $G_0(2)$ consists of matrices of
the form (\ref{sp}) with  $C = 2 C'$ for some integral $C'$.
Defining $B'=2B$ we see that under $G_0(2)$ transformation the theta
function transforms as
\begin{equation}\label{g2}
   \Theta_{E8}(2 \Omega) \rightarrow
    \Theta_{E8}(2 \frac{A \Omega+B}{C \Omega+D})=\Theta_{E8}( \frac{A
2\Omega+B'}{C' 2\Omega+D})
\end{equation}
Since $\Theta_{E8}(\Omega)$ is modular under $SP(2, \mathbb{Z})$, it
follows that $\Theta_{E_8}( \frac{A 2\Omega+B'}{C' 2\Omega+D})$ can
be reexpressed in terms of $\Theta_{E_8}(2 \Omega)$. We then
conclude that $\Theta_{E_8}(2 \Omega)$ is modular under $G_0(2)$.
Now, the sum of the partition functions for all twisted sectors is
modular invariant under $Sp(2,Z)$. The theta function $\Theta_{E8}(2
\Omega)$ will appear in several sectors, as $Z_{0001}$ or
$Z_{0010}$, etc. By inspection, one can also see that it always
appears accompanied by $\Phi_6(\Omega)$ in the denominator. The sum
of all partition functions will then be a sum of various modular
images of $\Theta_{E8}(2\rho, 2 \sigma, 2 v )$ over $\Phi_6$ as in
(\ref{twistefinal}). For this to be invariant under
$Sp(2,\mathbb{Z})$, and since the theta function in the numerator is
modular under $G_0(2)$, the denominators must be $G_0(2)$ modular
forms as well. Hence $\Phi_6(\Omega)$ must a modular form for
$G_0(2)$. This derivation also makes clear why the larger $G_0(2)$
appears even though the subgroup $G_1(2)$ is adequate for
accommodating the S-duality group $\Gamma_1(2)$.

Second, the precise cancelations of the Prym parameters that made
this result possible appear mysterious at first sight. However this
is not an accident but rather a consequence of the equivalence of
the $E_8 \times E_8$ and of the $(E_8 \times E_8)/{\mathbb{Z}_2}$
CFTs. This fact suggests another way to derive the required Siegel
modular forms which will be outlined in appendix \ref{B}.

\section{Generalization to $\mathbb{Z}_N$ Orbifolds \label{ZN}}

For higher $\mathbb{Z}_N$ orbifolds, the explicit CFT computations
are more complicated. For instance, the left-moving twist symmetry
$\xi$ is more involved for general $N$. However, using our
experience with the $\mathbb{Z}_2$ orbifold, we can streamline our
derivation and point out some of the essential ingredients in the
general $\mathbb{Z}_N$ case.

The dyon partition function of a $\mathbb{Z}_N$ CHL orbifold is
given in terms of the Siegel modular form $\Phi_k$. These forms are
defined by their modular properties and behavior at the boundary of
moduli space. No expression in terms of theta functions is available
but a CFT derivation along the lines outlined here will result in
such an expression. For now, one  proceed indirectly and argue that
the genus two chiral partition function for the CHL orbifold
involves a modular form with the same modular properties and value
at the boundary of moduli space as $\Phi_k$. There is no uniqueness
theorem proven for these forms however it is likely that they are
unique. Note that the form $\Phi_k$ is obtained by an additive lift
of the form $f_k(\tau)= \eta^{k+2}(\tau) \eta^{k+2}(N\tau)$ which is
known to be the unique cusp form of $\Gamma_1(N)$. We will also
argue that $\Phi_k$ have a nice expression in terms of theta
functions.

Let us recall some facts about a $\mathbb{Z}_N$ CHL orbifold. The
orbifold is defined in terms of a $\mathbb{Z}_N$ action on $N$
groups of $l$ left-moving bosons. If we diagonalize the
$\mathbb{Z}_N$ action then it is clear that $l$ bosons are left
invariant and the remaining  $l(N-1)$ bosons transform nontrivially.
As a result, $l(N-1)$ left-moving gauge fields are projected out and
the rank is reduced by $m =l(N-1) $ units. Starting with  rank $28$
for the toroidally compactified case, we thus obtain a model with a
reduced rank $r = 28 - l(N-1)$. {}From the list of known CHL models
for $N =2, 3, 5, 7$ to four dimensions that have $r=20, 16, 12, 10$
we conclude that $l = 8, 6, 4, 3$. We would now like to see how the
weight $k$ of the relevant Siegel modular form can be derived from
this data using the underlying picture of genus-two worldsheet.

To implement such a $\mathbb{Z}_N$ symmetry, one would have to start
with a self-dual Narain lattice $\Lambda$ of the form $\Gamma^{(lN,
lN-16)}$ which contains a sublattice $\Lambda_0$ of the form
$\Gamma^{(l , lN -16)}$ fixed by the $\mathbb{Z}_N$ action and an
orthogonal left-moving complement $\Lambda_1$ of the form
$\Gamma^{(l(N-1), 0)}$ that is rotated by the left-moving
$\mathbb{Z}_N$ action. Such a choice reduces the rank by $l (N-1)$
because gauge fields associated with the $\Lambda_1$ lattice are
eliminated. The lattice $\Lambda_1$ is strongly constrained by the
requirement of modular invariance and $\mathbb{Z}_N$ action.

The  computation of the twisted partition function proceeds as in
the $\mathbb{Z}_2$ case. The quantum piece of the $\mathbb{Z}_N$
orbifold twisted determinant can be computed from properties of
$SU(N)$ WZW model at level one by using the fact that $\mathbb{Z}_N$
orbifold symmetry can be viewed as a Weyl symmetry of the $SU(N)$.
Thus, by  $SU(N)$ conjugation, an order $N$ twist can again be
viewed as an order $N$ shift along the lattice. Using this trick,
the ratio between twisted and untwisted determinants can be written
again in terms of theta functions at genus one. At genus two it can
be written in $N$ different ways as a ratio between a twisted
$SU(N)$ theta function of the modular matrix and a $SU(N)$ theta
function of the Prym period.

Now, we have seen that the dependence on the Prym period cancels out
in the $\mathbb{Z}_2$ case in an apparently magical fashion from the
final expression for the partition function. This cancelation was
essential to obtain a nice Siegel modular form in the final answer
and was not an accident. In  the $\mathbb{Z}_2$ case, it is a
consequence of the fact that orbifolding by $\xi$ alone without
accompanying it with the shift $T$ along the circle leads back to
the original theory leading to the identity (\ref{e8identity}) and
its generalization to higher genera. We expect this to be true even
in the $\mathbb{Z}_N$. Otherwise, it would lead to an undesirable
dependence on the Prym periods. Such a computation would provide a
useful representation of these forms $\Phi_k$ in terms of theta
functions.

With this information we can deduce the weight of the relevant
Siegel modular form quite easily. Basically, the point is that the
twisted sector partition function  is short of a lattice sum over
$\Lambda_1$ in the numerator. A momentum sum over the $m$
dimensional lattice would have led  to a theta function with modular
weight $m/2$ which would now be missing. To compensate for this, the
twisted determinants in the denominator should also have a
correspondingly smaller weight. Putting all factors together  we get
the modular form $\Phi_{10}$ of weight $10$ in the denominator of
the untwisted partition sum. Hence we expect that the modular form
in the denominator in the twisted sector will  have weight $k =
(10-m/2)$. Substituting $l(N+1) = 24$ and $m = l(N-1)$, we obtain
$k= (l-2)$ or $k = 24/(N+1) -2$  precisely in agreement with the
proposed relation (\ref{relation}). Note that in
\cite{Jatkar:2005bh}, the relation between $k$ and $N$ was put in by
hand to obtain agreement with the subleading terms in the
Bekenstein-Hawking-Wald entropy. Here we are able to derive it from
general considerations of the genus-two partition function of the
orbifold.  We have summarized various parameters of the orbifolds in
the table (\ref{relation}).
\begin{table}
  \centering
  \begin{tabular}{|c|c|c|c|c|c|c|c|}
  \hline
  N & l  & k & r & m & $\Lambda_0$ & $\Lambda_1$ & $\Lambda$ \\ \hline
  2 & 8 & 6 & 20 & 8 & $\Gamma^{(8, 0)}$ & $\Gamma^{(8, 0)}$ & $\Gamma^{(16, 0)}$\\
  3 & 6 & 4 & 16 & 12 & $\Gamma^{(6,2)}$ & $\Gamma^{(12, 0)}$& $\Gamma^{(18, 2)}$ \\
  5 & 4 & 2 & 12 & 16 &$\Gamma^{(4, 4)}$ & $\Gamma^{(16, 0)}$& $\Gamma^{(20, 4)}$\\
  7 & 3 & 1 & 10 & 18 & $\Gamma^{(3, 5)}$ & $\Gamma^{(18, 0)}$ & $\Gamma^{(21, 5)}$\\
  \hline
\end{tabular}
  \caption{List of parameters for available CHL Orbifolds}\label{table}
\end{table}
Note that the $\mathbb{Z}_7$ orbifold is qualitatively different
since it requires a $(21, 5)$ lattice and thus one does not a fully
factorized $ K3 \times T^2$ that was necessary to use the string web
picture. It is possible therefore that this case requires a slightly
different treatment.

\section{Comments \label{Comments}}

We conclude with a brief comment on the fermionic zero modes. We
have not dealt with the right-moving superstring carefully.
Effectively, our prescription was to evaluate the genus two
partition function of the bosonic string and then use holomorphic
factorization to read off the left-moving part. It should be
possible to define an intrinsically superstring amplitude. For
example, in the Green-Schwarz formalism, the full partition function
vanishes because of the fermion zero modes but an insertion of an
appropriate number of  fermionic currents can soak up the
right-moving fermion zero modes. The nonzero modes of the
Green-Schwarz fermions and the light-cone bosons are expected to
cancel in pairs in the right-moving partition function leaving
behind only the left-moving partition function. This would
correspond to an appropriate index-like quantity such as a helicity
supertrace. Such a prescription surely works at genus-one to
correctly obtain the helicity supertrace that counts the heterotic
half-BPS states. At genus two, the Green-Schwarz superstring is more
subtle but it would be desirable to have  an appropriate definition
of index-like quantity that is non-vanishing after soaking up the
zero modes and counts the left-moving fluctuations.

The genus-two picture outlined here is easily amenable to
generalizations to other more general orbifolds \cite{David:2006ru,
David:2006ud}. These  generalizations  and an elaboration of the
computation outlined in appendix \eqref{B} will be explored
elsewhere. The genus two picture also raises the question of
possible contributions of higher genus surfaces. This question will
be addressed in the forthcoming publication \cite{Dabholkar:2006mm}.

\subsection*{Acknowledgements}

We would like to thank Ashoke Sen and Herman Verlinde for useful
discussions. A. D. would like to thank Harvard University, LPTHE at
the University of Paris, and Aspen Center for Physics for
hospitality where part of this work was completed. D. G. would like
to thank TIFR for hospitality where this work was initiated.

\appendix

\section{Computation of the Classical Piece of the Partition Function}

The classical piece coming from the sum over momenta involving the
period matrix $\Omega$ and the Prym period $\tilde \tau$ is best
understood in terms of the period matrix of the covering space $\hat
\Sigma$ of genus-three.  The genus-three theta function for an $E_8$
factor (\ref{momsumhat}) involves a sum over momenta ${\bf p} =
(p_1, p_2, p_3)$ where each momentum $p_i$ takes values in the $E_8$
root lattice $\Lambda$ with length-squared two. Using the specific
form of the period matrix for $\hat \Sigma$ in (\ref{sigmahat}) we
can write it in terms of genus-two objects as
\begin{equation}\label{twosum}
\sum_{p_1,p_2,p_3 \in \Lambda}
    \exp\left[ \pi i
    {\frac{p_1 + p_3}{\sqrt{2}} \choose \sqrt{2} p_2}
    \cdot \Omega \cdot {\frac{p_1 + p_3}{\sqrt{2}} \choose \sqrt{2} p_2} \right]
   \times \exp\left[ \pi i \frac{p_1 - p_3}{\sqrt{2}} \cdot \tilde\tau \cdot
   \frac{p_1 - p_3}
    {\sqrt{2}} \right] \,\, .
\end{equation}
The vector $\sqrt{2} p_2$ lies in the ``level  two'' lattice
$\Lambda[2]$ which is a lattice with twice as large length-squared
compared to $\Lambda$. The vectors $({p_1 + p_3})/\sqrt{2}$ are in a
copy of $\Lambda[2]$ shifted by some element ${\cal P}$ in the coset
$\Lambda[2]^*/\Lambda[2]$, $({p_1 - p_3})/\sqrt{2}$ in another copy
of $\Lambda[2]$ shifted by the same ${\cal P}$. In the specific
example of $\Lambda$ being the lattice of $E_8$, there are $256$
values of ${\cal P}$, collected into orbits of the Weyl group of the
fundamental weight of the trivial, adjoint, and 3875
representations, of lengths 1, 120, and 135, respectively
\cite{Dabholkar:2005dt}. This allows one to reorganize the
expression above as
\begin{equation}
\sum_{{\cal P}\in \frac{\Lambda}{2\Lambda}} \left( \sum_{q_1 \in
\Lambda[2] \atop q_2 \in \Lambda[2]+{\cal P}}
    \exp\left[\pi i {q_1 \choose q_2} \cdot \Omega \cdot {q_1 \choose q_2} \right]
    \times \sum_{q_3 \in \Lambda[2]+{\cal P}}
    \exp\left[\pi i q_3 \cdot \tilde\tau \cdot q_3 \right] \right),
\end{equation}
The momentum sum involving the Prym period can be readily performed.
It is useful to  introduce $E_8$ theta functions with
characteristics:
\begin{equation} \label{thp}
\Theta_{E_8[2], \wp}({\tilde\tau}) := \sum_{\Delta\in E_8(1)}
e^{2\pi i {\tilde\tau}( \Delta - \frac12 \wp)^2 }
\end{equation}
The theta series  are thus the numerators of affine characters of
$E_8$ at level 2, and can be computed explicitly using free fermion
representations or by inspection in the bosonic representation
\cite{Dabholkar:2005dt}:
\begin{equation}
\begin{array}{lcccc} \label{thetas}
\Theta_{E_8[2],1}({\tilde\tau}) &=& \Theta_{E_8[1]}(2{\tilde\tau})
&=& \frac{1}{16}(\vartheta_3^4 \vartheta_2^4 - \vartheta_4^4
\vartheta_2^4+16 \vartheta_3^4 \vartheta_4^4) \\
\Theta_{E_8[2],248} ({\tilde\tau}) &=& \frac12 \left( \th_3^6
\th_2^2 + \th_2^6 \th_3^2 \right)(2{\tilde\tau}) &=&
\frac{1}{16}(\vartheta_3^4 \vartheta_2^4 + \vartheta_4^4
\vartheta_2^4) \\
\Theta_{E_8[2],3875}({\tilde\tau}) &=& \th_3^4 \th_2^4
(2{\tilde\tau}) &=& \frac{1}{16}(\vartheta_3^4 \vartheta_2^4 -
\vartheta_4^4 \vartheta_2^4)
\end{array}
\end{equation}

Note that the combination $\vartheta_3^4 \vartheta_4^4$ appears only
in $\Theta_{E_8[2],1}$ corresponding to ${\cal P} =0$. Hence the
coefficient of $\vartheta_3^4 \vartheta_4^4$ is just
\begin{equation} \sum_{q_1 \in \Lambda[2]\atop q_2
\in \Lambda[2]} \exp\left[\pi i {q_1 \choose q_2} \cdot \Omega \cdot
{q_1 \choose q_2}\right] = \Theta_{E8}(2 \rho,  2 \sigma, 2v).
\end{equation}
Similarly,  the combination $\vartheta_3^4 \vartheta_2^4$ appears
for all three classes of $\cal P$ in \eqref{thetas} with the same
coefficient. Thus, in  \eqref{twosum} we now have the sum over $q_2$
taking values in $\Lambda[2] + \cal P$  with an unrestricted sum
over $\cal P$. We can thus replace these two sum by a single sum
over $q_2$ that takes values in $\Lambda^*$. As a result, the
coefficient of $\vartheta_3^4 \vartheta_2^4$ is simply
\begin{equation}
\frac{1}{16} \sum_{q_1 \in \Lambda[2] \atop q_2 \in \Lambda[2]^*}
\exp\left[\pi i {q_1 \choose q_2} \cdot \Omega \cdot {q_1 \choose
q_2} \right] = \frac{1}{16} \Theta_{E8}(\frac{\rho}{2},  2 \sigma,
v).
\end{equation}  By a similar reasoning, the coefficient of
$\vartheta_4^4 \vartheta_2^4$ is
\begin{equation}
-\frac{1}{16} \sum_{q_1 \in \Lambda[2] \atop q_2 \in \Lambda[2]^*}
(-1)^{q_2^2}\exp\left[\pi i {q_1 \choose q_2} \cdot \Omega \cdot
{q_1 \choose q_2}\right] = -\frac{1}{16}
\Theta_{E8}(\frac{\rho+1}{2}, 2\sigma,  v).
\end{equation}
Here the term $(-1)^{q_2^2}$ accounts for the fact that the $248$
representation has relative minus sign with respect to the other two
representations for the coefficient of $\vartheta_4^4 \vartheta_2^4$
in \eqref{thetas}. Putting it together we see that the momentum sum
\eqref{momsumhat} equals
\begin{eqnarray}\label{identity}
\nonumber \Theta_{E8}(2\rho,  2\sigma, 2v
)\vartheta^4_{00}({\tilde\tau}) \vartheta^4_{01}({\tilde\tau}) +
\frac{1}{16}\Theta_{E8}(\rho/2, 2\sigma, v
)\vartheta^4_{00}({\tilde\tau}) \vartheta^4_{10}({\tilde\tau}) -
\frac{1}{16}\Theta_{E8}(\frac{\rho+1}{2}, 2\sigma, v
)\vartheta^4_{01}({\tilde\tau}) \vartheta^4_{10}({\tilde\tau}).
\end{eqnarray}

\section{An Outline of An Alternative Derivation \label{B}}

It is well-known that if we take the orbifold generator to be $\xi$
alone without the half-shift $T$ along the circle, then the
resulting orbifold gives back the same theory. The twisted sectors
have integral conformal dimensions, and neatly take the place of the
odd $E_8$ currents that are removed by the orbifolding projection.
This equivalence suggests another way to derive the expression for
$\Phi_6$ in terms of $\Phi_{10}$ and theta function.

For simplicity, let us first consider the genus-one case. The above
equivalence implies in particular the equality of the torus
partition function of the two theories
\begin{equation}
\frac{\Theta^2_{E8}({\tau})}{\eta^{16}({\tau})} =
\frac{1}{2}\frac{\Theta^2_{E8}({\tau})}{\eta^{16}({\tau})} +
\frac{1}{2}\frac{\Theta_{E8}(2 {\tau})}{\eta^8(2 {\tau})} +
\frac{1}{2}\frac{\Theta_{E8}(\frac{{\tau}}{2})}{\eta^8(\frac{{\tau}}{2})}+
\frac{1}{2}\frac{\Theta_{E8}(\frac{{\tau}+1}{2})}{\eta^8(\frac{{\tau}+1}{2})}
\end{equation}
By some simple $\vartheta$ identities this can be rewritten as a
useful theta function equality
\begin{equation}\label{e8identity}
\Theta^2_{E8}(\tau) = \Theta_{E8}(2 \tau)\vartheta^4_{00}(\tau)
\vartheta^4_{10}(\tau) +
\Theta_{E8}(\frac{{\tau}}{2})\vartheta^4_{00}({\tau})
\vartheta^4_{10}({\tau}) -
\Theta_{E8}(\frac{{\tau}+1}{2})\vartheta^4_{01}({\tau})
\vartheta^4_{10}({\tau}).
\end{equation}
This identify implies
\begin{equation}\label{oneloop}
    \frac{\Theta^2_{E8}(\tau)}{\eta^{24}(\tau)} = \frac{\Theta_{E8}(2
    \tau)}{\eta^8(\tau)\eta^8(2\tau)} + \ldots
\end{equation}

Note that $\eta^{24}$ is the genus-one analog of $\Phi_{10}$ whereas
$\eta^8(\tau)\eta^8(2\tau)$ is the genus-one analog of $\Phi_6$
associated with the counting of electric states.  We therefore
expect a similar identity for the genus two partition function which
follows from the above equivalence of orbifolded and unorbifolded
theory. It should read
\begin{equation}
\Theta^2_{E8}(\Omega) =
\Theta_{E8}(2\Omega)\vartheta_{0000}^2(\Omega)\vartheta_{0001}^2(\Omega)
\vartheta_{0010}^2(\Omega)\vartheta_{0011}^2(\Omega) + \cdots
\end{equation}
\begin{equation}
\frac{\Theta^2_{E8}(\Omega)}{\Phi_{10}(\Omega)} =
\frac{\Theta_{E8}(2\Omega)}{\Phi_6(\Omega)} + \cdots
\end{equation}
This expression has to coincide with the sum of all the twisted
partition functions $Z_{abcd}$. It clearly does not involve any
explicit dependence on the Prym parameters, and indicates that the
dependence should drop off the single $Z_{abcd}$ as well. It also
motivates more directly the expression of $\Phi_6$ in terms of theta
functions. We expect that  these ideas can be generalized to the
general $\mathbb{Z}_N$ orbifolds.

\bibliographystyle{JHEP}
\bibliography{web}

\end{document}